\documentclass[12pt, draftclsnofoot,onecolumn]{IEEEtran}
\usepackage[usenames]{color}
\usepackage{amsfonts}
\usepackage{cite}
\usepackage{amssymb,amsthm}
\usepackage{bbm, amsmath}
\usepackage[caption=false]{subfig}
\usepackage{graphicx}
\usepackage{footnote}
\usepackage{algpseudocode}
\usepackage{algorithm}
\usepackage{multirow}
\usepackage{amsmath}
\usepackage{xcolor}
\usepackage{verbatim}
\usepackage{times}
\usepackage{epstopdf}
\usepackage{bm}
\usepackage{color,soul}
\usepackage{tikz}

\usetikzlibrary{patterns}

\usepackage{pifont}

\title{Performance of Energy Harvesting Receivers with Power Optimization}
\author{Zhengwei~Ni, Mehul~Motani
	\thanks{The authors are with Department of Electrical \& Computer Engineering, National University of Singapore, Singapore (email:
		a0123808@u.nus.edu, motani@nus.edu.sg). This work was supported in part by the National Research Foundation Singapore under Grant No. NRF-CRP-8-2011-01. Some of the results in this paper have been submitted to IEEE Globecom 2017.}
		}

\date{\today}

\begin{document}
\maketitle

\begin{abstract}
The difficulty of modeling energy consumption in communication systems leads to challenges in energy harvesting (EH) systems, in which nodes scavenge energy from their environment.  An EH receiver must harvest enough energy for demodulating and decoding. The energy required depends upon factors, like code rate and signal-to-noise ratio, which can be adjusted dynamically.  We consider a receiver which harvests energy from ambient sources and the transmitter, meaning the received signal is used for both EH and information decoding.  Assuming a generalized function for energy consumption, we maximize the total number of information bits decoded, under both average and peak power constraints at the transmitter, by carefully optimizing the power used for EH, power used for information transmission, fraction of time for EH, and code rate. For transmission over a single block, we find there exist problem parameters for which either maximizing power for information transmission or maximizing power for EH is optimal.  In the general case, the optimal solution is a tradeoff of the two. For transmission over multiple blocks, we give an upper bound on performance and give sufficient and necessary conditions to achieve this bound. Finally, we give some numerical results to illustrate our results and analysis.

\end{abstract}
\newtheorem{mydef1}{Lemma}
\newtheorem{mydef2}{Theorem}
\newtheorem{mydef3}{Remark}
\newtheorem{mydef4}{Definition}
\newtheorem{mydef5}{Corollary}
\section{Introduction}
Energy harvesting techniques enlarge the mobility of devices by breaking away from the limitations of the conventional power supplies, and give the freedom to deploy networks at hard-to-reach places, such as remote areas and the human body. As such, energy harvesting networks have the potential to be implemented in many new areas including medical, environmental and safety applications.

Dependent on the characteristics of energy sources, many energy harvesting techniques are under investigation. Among them, radio frequency (RF) radiation is a promising technique and has already been used in many applications. The survey paper \cite{SurveyApplication} offers some examples, such as Computational RFID \cite{4539485}. Additionally, researchers at the University of Washington have deployed an energy scavenging WiFi camera \cite{7113088}. Specially, the transmitter can work as an energy source, as in the dedicated RF charging \cite{RPercy,RMascarenas}. Thus, both energy and information can be delivered to the receiver via RF waves \cite{RLav,RAnshoo}.  Compared with ambient energy sources, the amount of energy harvested from dedicated RF sources can be controlled and dynamically adjusted.

As we know, a receiver must harvest enough energy for demodulating and decoding, and for many systems, the energy consumed at the receiver can be comparable to or even larger than that at the transmitter \cite{karl2007protocols}. However, the difficulty of modeling energy consumption at the receiver leads to challenges. Some papers on energy harvesting ignore it \cite{RGurakan1,RGurakan2} or assume it is given as a constant \cite{RLav,RAnshoo}. The relationships between energy required at the receiver and other parameters or performance metrics are not clear.


In our previous works \cite{RNi1,RNi2}, we study the case that the transmitter works as a dedicated energy source and provide some energy to the receiver, which has a time-switching architecture \cite{RSurvey}. To address the problem mentioned above, we make the following assumptions. Firstly, the authors of \cite{RGrover} find that one can focus on energy consumed for decoding because they observe that allowing uncoded transmission significantly reduces the system energy consumption in designing transceiver circuits. Motived by this work, we assume the energy consumption of other processing functions for extracting information is negligible compared with decoding. In addition, energy consumption at the decoder is highly dependent on the decoding scheme used and hardware implementation. To find a bridge to connect energy consumption and the performance metrics we are interested in, we follow the approach in \cite{RDoost1,RDoost2,RDoost3} that uses a generalized function to express the energy consumed at the decoder in terms of code rate and channel capacity. Based on these settings, by carefully allocating the time for energy harvesting/information transmission and choosing the code rate, we maximize the total amount of information decoded for various scenarios.

In these previous works, we assume the symbols are transmitted at a predetermined constant power. The channel is unchanging and we do not consider possible manipulation of the power at the transmitter side. However, improving the performance by increasing the signal-to-noise ratio is a fundamental technique in wireless communication systems. We can also control the amount of energy transferred by adjusting the power. Hence, in this paper, we extend our previous work \cite{RNi1,RNi2} by allowing optimization of the powers used for both information transmission and energy harvesting, in addition to optimizing the fraction of time for energy harvesting and the code rate.  Intuitively, to maximize the number of information bits decoded at the receiver, the transmitter has two choices on how to allocate its transmit power:
\begin{enumerate}
	\item To increase the transmitting power for information transmission. In this scenario, the signal-to-noise ratio increases and the symbol error rate will decrease. Thus, the channel condition becomes better.\footnote{For example, if Binary Phase Shift Keying (BPSK) is used for modulation and receiver use hard decision decoding, the channel can be modeled as a binary symmetric channel (BSC). When transmitting power increases, the crossover probability decreases, meaning that the channel capacity becomes larger.} Hence, it will cost less energy to extract the information, and the transmitter may transfer less energy to the receiver from the dedicated energy signals.
	\item To increase the transmitting power for energy harvesting. In this scenario, the transmitter tries to transfer energy to the receiver more efficiently, so more time can be used for information transmission even though the channel conditions may not be so good.
\end{enumerate} 
Then, the basic question is:  \textbf{For given power constraints at the transmitter, how should the transmitter allocate power based on the two scenarios described above?}

Motivated by this question, in this paper, we consider an end-to-end communication system, in which the receiver can harvest energy from the environment. In addition, the transmitter also offers energy to the receiver for powering its circuitry. Both average and peak power constraints are considered at the transmitter. The aim is to maximize the total number of information bits decoded at the receiver by joint optimization of the power used for energy harvesting, power used for information transmission, fraction of time for energy harvesting, and code rate, over both single block or multiple blocks. The contributions of our paper are summarized as follows:
\begin{enumerate}
	\item For transmission over a single block, we formulate a non-convex optimization problem to maximize the amount of information decoded at the receiver. Then, based on our observation about the structure of the problem, we find a method to obtain all local optimal solutions by solving a series of equations. These local optimal solutions correspond to three different schemes: maximizing power used for energy harvesting, maximizing power used for information transmission, and a tradeoff of the two as a general case. Our results show that these three schemes all have the potential to be optimal. Finally, the global optimal solution can be determined by comparing these local optimal solutions.
	\item For transmission over multiple blocks, we give an upper bound on the total number of bits decoded at the receiver. Furthermore, we also provide sufficient and necessary condition to achieve this upper bound. For other cases, where the condition cannot be satisfied, we obtain the local optimal solution by an iterative algorithm.
	\item Finally, in the numerical results, we consider an example that coincides with low-density parity-check (LDPC) codes. In this example, we find that we should maximize the transmitting power for information transmission when there is a relatively strict constraint on peak power. Maximizing the transmitting power for energy harvesting is the best scheme when both average power constraint and peak power constraint are loose. If it is neither of these two cases, we should make a trade-off between them. 
	
\end{enumerate}

\section{System Model}
In this paper, we consider an end-to-end communication system with an energy harvesting receiver. The receiver receives signals transmitted by the transmitter and tries to extract the information contained inside by demodulation and decoding. To support powering the circuitry for receiving and processing, the receiver also needs to harvest energy from outside. A time-switching architecture is designed for the receiver in the context of energy harvesting. In this architecture, a switcher inside the receiver can connect the RF front-end circuits to either energy harvesting or information receiving sub-systems. Thus, the RF signals radiated by the transmitter can be used for either harvesting energy or extracting information.

The duration of one block is $T$. In a given block, a scheme called Harvest-then-Receive (HTR) is used \cite{RNi1}. HTR operates as follows: for the first $\alpha T$ duration, the switcher connects the RF front-end to the energy harvesting sub-system. At the same time, the transmitter transmits RF signals which contain no information but are specially designed for energy harvesting. Then for the reminder $(1-\alpha)T$ duration, the switcher connects the RF front-end to the information receiving sub-system and the transmitter starts information transmission. We call $\alpha$ the fraction of time for energy harvesting.

Now, we present the channel model for information transmission. An information sequence is encoded using capacity approaching/achieving channel codes. This code has a binary input alphabet, which coincides with many popular channel codes, such as LDPC and polar codes. Then, the encoded information is modulated using BPSK. The modulated symbols are sent through an additive white Gaussian noise (AWGN) channel with power spectral density (PSD) $N_0/2$. Without loss of generality, we set $N_0=1$. Letting $p^{\rm I}$ be the average power per symbol and $T_{\rm S}$ be the symbol duration, the symbol (bit) error rate is $Q(\sqrt{2e^{\rm I}})$, where $Q(x)=\int_{x}^{+\infty}\frac{1}{\sqrt{2\pi}}e^{-\frac{t^2}{2}}dt$ and $e^{\rm I}=p^{\rm I}T_{\rm S}$ is energy per symbol. The receiver performs a hard decision on the received symbols then starts channel decoding, so the channel can be regarded as a BSC with crossover probability equivalent to the symbol (bit) error rate $Q(\sqrt{2e^{\rm I}})$. Here we want to emphasize that even though we consider BPSK, it is easy to extend the results to other modulation schemes using some well-known approximate bit error probabilities \cite[Table 6.1]{goldsmith2005wireless}. Note that since the duration of one block is $T$, one block can be discretized into $n=T/T_{\rm S}$ channel uses, where each symbol is sent using one channel use.

For energy harvesting, the receiver can harvest energy from both the transmitter and other ambient sources. Here we do not limit the other sources to be RF and allow for solar, wind, etc. Assuming the transmitting power for energy harvesting is $p^{\rm E}$,  it can be regarded as $\alpha T/T_{\rm S}=\alpha n$ channel uses are used for energy harvesting and the energy transmitted per channel use is $e^{\rm E}=p^{\rm E}T_{\rm S}$. At the receiver, the energy can be obtained via RF to DC conversion. However, this conversion depends on many factors, such as rectenna and impedance matching between the antenna and the voltage multiplier, and a certain amount of energy may be lost during conversion \cite{RRectenna1,RRectenna2}. Hence, we set the conversion efficiency to be $\eta$, where $0<\eta\leq1$. 

We follow our previous work \cite{RNi1,RNi2} to model the energy required for extracting information at the receiver. As previously mentioned, firstly, we assume the energy used at other components for extracting information is negligible. Then, instead of giving an exact expression for how much energy is consumed, we use a generalized function, $\mathcal{E}_{\rm D}(\theta)$, to express the energy consumed for decoding per channel use. $\theta$ is the inverse of capacity gap, which is defined as $\delta=1-R/C$, where $R$ is code rate and $C$ is channel capacity. That is, $\theta = 1/\delta = C/(C-R)$. Furthermore, we require this generalized function to satisfy certain properties as follows:
\begin{enumerate}
	\item[(1)] $\mathcal{E}_{\rm D}(1)=0$. When $\theta\rightarrow +\infty$, $\mathcal{E}_{\rm D}(\theta)\rightarrow +\infty$.
	\item [(2)] $\mathcal{E}_{\rm D}(\theta)$ is a non-decreasing convex function of $\theta$.
\end{enumerate}
Briefly, the reason that we can choose this function and these properties to characterize the energy consumed at the decoder is that the capacity gap is widely used in the research on the decoding complexity of capacity approaching/achieving codes based on iterative decoding, and the given properties coincide with their results \cite{RDoost2,RLDPC1,RLDPC2} if we assume the energy consumed for decoding is proportional to the complexity of decoding scheme. A detailed elaboration is given in our previous work \cite{RNi1}. 
The decoder starts working after all the symbols have been received, which is at the end of a block, so the energy harvested can be used to decode symbols received in the same block.

In addition, extra energy is consumed for processing and analyzing of the received data. We assume the amount of energy used for these operations at receiver is $\tilde{G}$ and it is also used at the end of one block, after all the information is decoded. Since all of these operations are controlled by the processor at the receiver, the energy consumed for this part can be well predicted in the short term, which makes offline optimization possible. We use $\tilde{g}=\tilde{G}/n$ to express the energy consumed for other operations per channel use. The energy received from other sources may vary with time. However, since decoding starts after energy harvesting, we only care about the average, which is defined as $e^{{\rm ambient}}$ per channel use in a similar way. To make dedicated RF charging meaningful, we assume the energy harvested from other sources is far from enough. Thus, $g = \tilde{g}- e^{{\rm ambient}} \geq 0$.

Both average and peak power constraints are considered at the transmitter. The average transmitting power should not be larger than $p^{\rm avg}$, meaning that $\alpha p^{\rm E}T+(1-\alpha) p^{\rm I}T\leq p^{\rm avg}T$. We define $e^{\rm avg}=p^{\rm avg}T_{\rm S}$, so the previous inequality is equivalent to $\alpha e^{\rm E} + (1-\alpha) e^{\rm I} \leq e^{\rm avg}$. We also set $p^{\rm I}\leq p^{\rm lim}$ and $p^{\rm E}\leq p^{\rm lim}$ for the peak power constraint, or equivalently $e^{\rm I}\leq e^{\rm lim}$ and $e^{\rm E}\leq e^{\rm lim}$, where $e^{\rm lim}=p^{\rm lim}T_{\rm S}$. We assume $\eta e^{\rm avg}-g \geq 0$ so there does not exist the case that even if all energy is used for energy harvesting, it still cannot support the consumption at the receiver. In addition, to make the peak power constraints meaningful, we assume $e^{\rm avg} < e^{\rm lim}$.

We consider transmission over both single and multiple blocks. For multiple blocks, we assume the information is delay-sensitive, meaning that the decoder is not allowed to store received symbols and decode them in the following blocks when it has harvested enough energy. However, the energy harvested in one block can be stored in a battery and used in the future. We do not limit the size of the battery so there will be no energy wasted due to overflow.

\section{Transmission over a Single Block}\label{singleblock}
In this section, we investigate the performance over a single block. The transmitter and the receiver want to maximize the number of bits decoded by adjusting the fraction of time (channel uses) for energy harvesting $\alpha$, code rate $R$, energy used for information transmission $e^{\rm I}$ and energy harvesting $e^{\rm E}$ per channel use. The optimization problem can be given as
\\* (P1)
\begin{eqnarray}
	\max\limits_{\alpha, R, e^{\rm E}, e^{\rm I}}&\quad&(1-\alpha)R,\\{\rm s.t.}&\quad&(1-\alpha)\mathcal{E}_{\rm D}(\theta)+g\leq \eta\alpha e^{\rm E},
	\label{P1_first_contraints}\\{}&\quad& \alpha e^{\rm E} + (1-\alpha) e^{\rm I} \leq e^{\rm avg}, \label{P1_second_contraints}\\{}&\quad&\theta = \frac{C(e^{\rm I})}{C(e^{\rm I})-R},\label{P1_theta_constraints}\\{}&\quad& 0\leq e^{\rm E}\leq e^{\rm lim},\label{P1_ee_contraints}
	\\{}&\quad& 0\leq e^{\rm I}\leq e^{\rm lim}\label{P1_ii_contraints},
	\\{}&\quad& 0\leq \alpha\leq 1,\label{P1_alpha_contraints}
	\\{}&\quad& 0\leq R\leq C(e^{\rm I}) \label{P1_last_contraints}.
\end{eqnarray}
The total number of bits decoded at receiver is $(1-\alpha)Rn$, here we only maximize $(1-\alpha)R$ because the number of channel uses in one block $n$ is a constant. \eqref{P1_first_contraints} comes from energy causality, meaning that the energy consumed at receiver should not be larger than that harvested. \eqref{P1_second_contraints} is from average power constraint while \eqref{P1_ee_contraints} and \eqref{P1_ii_contraints} are from peak power constraint. In general, the channel capacity is a function of $e^{\rm I}$. As we mentioned in the previous section, the channel is a BSC with crossover probability equal to $Q\big(\sqrt{2e^{\rm I}}\big)$, so the capacity is given as
\begin{eqnarray}
C(e^{\rm I})=1+Q\big(\sqrt{2e^{\rm I}}\big)\log_2 \big(Q\big(\sqrt{2e^{\rm I}}\big)\big)+\big(1-Q\big(\sqrt{2e^{\rm I}}\big)\big)\log_2\big(1-Q\big(\sqrt{2e^{\rm I}}\big)\big).	
\end{eqnarray}
To solve P1, we first give two useful lemmas.
\begin{mydef1}\label{lemma1}
To be optimal, \eqref{P1_first_contraints} must hold with equality.	
\end{mydef1}
\begin{proof}
	If the equality in \eqref{P1_first_contraints} does not hold, we can increase $\theta$ to make the equality hold since $\mathcal{E}_{\rm D}(\theta)$ is a non-decreasing function of $\theta$. According to \eqref{P1_theta_constraints}, when $e^{\rm I}$ is fixed, increasing $\theta$ means increasing $R$, so the value of objective function is also increasing. 
\end{proof}
\begin{mydef1}\label{lemma2}
	To be optimal, \eqref{P1_second_contraints} must hold with equality.
\end{mydef1}
\begin{proof}
	According to Lemma \ref{lemma1}, we can express $e^{\rm E}$ in terms of other parameters and substitute it into \eqref{P1_second_contraints}. Then, we can obtain
	\begin{eqnarray}
	\frac{(1-\alpha)\mathcal{E}_{\rm D}(\theta)+g}{\eta}+(1-\alpha)e^{\rm I}\leq e^{\rm avg}.\label{P1_take_first_constraints_into_second_constraints}
	\end{eqnarray}
	Similarly, we can decrease $\alpha$ in \eqref{P1_take_first_constraints_into_second_constraints} to increase the value of objective function if the equality does not hold.	
\end{proof}

\begin{mydef3}
Intuitively, Lemma \ref{lemma1} is true because, for the single block case, it is better to use up the energy harvested.
Intuitively, Lemma \ref{lemma2} is true is because we should use up the energy to transmit as much as possible to achieve better performance. 
\end{mydef3}

Then, based on Lemma \ref{lemma1}, Lemma \ref{lemma2} and \eqref{P1_theta_constraints}, 
	we can express $\alpha$, $R$ and $e^{\rm E}$ in terms of $\theta$ and $e^{\rm I}$, as
	\begin{eqnarray}
	\alpha &=& 1-\frac{\eta e^{\rm avg}-g}{\eta e^{\rm I}+\mathcal{E}_{\rm D}(\theta)},\label{expressing_alpha}\\
	R &=& \frac{\theta-1}{\theta}C(e^{\rm I}),\label{expressing_R}\\
	e^{\rm E}&=&\frac{\mathcal{E}_{\rm}(\theta) e^{\rm avg}+g e^{\rm I}}{\eta e^{\rm I}+\mathcal{E}_{\rm D}(\theta)-(\eta e^{\rm avg}-g)}\label{expressing_ee}.
	\end{eqnarray}
	Notice that, as mentioned in the previous section, we assume $\eta e^{\rm avg}-g \geq 0$, so $\alpha\leq 1$. Then, (P1) can be simplified into (P2), as 
	\\* (P2)
	\begin{eqnarray}
			\max\limits_{\theta, e^{\rm I}}&\quad&\frac{\theta-1}{\theta}\cdot (\eta e^{\rm avg}-g)\cdot\frac{C(e^{\rm I})}{\eta e^{\rm I}+\mathcal{E}_{\rm D}(\theta)}\label{P2O},\\{\rm s.t.}&\quad& 0\leq e^{\rm I}\leq e^{\rm lim},\label{ineq_constrain_1}
			\\{}&\quad& \theta \geq 1,\label{ineq_constrain_2} 
			\\{}&\quad& \mathcal{E}_{\rm D}(\theta)+\frac{\eta e^{\rm lim}-g}{e^{\rm lim}-e^{\rm avg}} e^{\rm I} \geq \frac{\eta e^{\rm avg}-g}{e^{\rm lim}-e^{\rm avg}} e^{\rm lim},\label{ineq_constrain_4}			
	\end{eqnarray}	
where \eqref{ineq_constrain_2} is obtained by substituting \eqref{expressing_R} into \eqref{P1_last_contraints}. In addition, \eqref{ineq_constrain_4} is obtained by substituting \eqref{expressing_alpha} into \eqref{P1_alpha_contraints} and substituting \eqref{expressing_ee} into \eqref{P1_ee_contraints}. In addition, \eqref{ineq_constrain_2} can be replaced by $\mathcal{E}_{\rm D}(\theta)\geq 0$ equivalently due to our definitions and assumptions in the previous section.

Even though we only need to optimize two parameters now, (P2) is still challenging to solve. Based on our observation of (P2), we can find that when we fixed one parameter and optimize the other one, the problem becomes to reveal unimodality. By using this property, for any fixed $e^{\rm I}>0$, we can find the corresponding $\theta$ that maximizes the objective function (P3), which is given as
\\* (P3)
	\begin{eqnarray}
		\max\limits_{\theta}&\quad&\frac{\theta-1}{\theta}\cdot (\eta e^{\rm avg}-g)\cdot\frac{C(e^{\rm I})}{\eta e^{\rm I}+\mathcal{E}_{\rm D}(\theta)},\label{P3O}\\{\rm s.t.}&\quad& \mathcal{E}_{\rm D}(\theta) \geq \max\bigg\{0,\frac{\eta e^{\rm avg}-g}{e^{\rm lim}-e^{\rm avg}} e^{\rm lim}-\frac{\eta e^{\rm lim}-g}{e^{\rm lim}-e^{\rm avg}} e^{\rm I}\bigg\} \label{P3C},	
	\end{eqnarray}
 Similarly, for any fixed $\theta>1$, we can find the corresponding  $e^{\rm I}$ that maximizes the objective function (P4), which is given as
\\* (P4)
 \begin{eqnarray}
 	\max\limits_{e^{\rm I}}&\quad&\frac{\theta-1}{\theta}\cdot (\eta e^{\rm avg}-g)\cdot\frac{C(e^{\rm I})}{\eta e^{\rm I}+\mathcal{E}_{\rm D}(\theta)},\label{P4O}\\{\rm s.t.}&\quad& \max\bigg\{0, \frac{\eta e^{\rm avg}-g}{\eta e^{\rm lim}-g}e^{\rm lim}-\frac{e^{\rm lim}-e^{\rm avg}}{\eta e^{\rm lim}-g}\mathcal{E}_{\rm D}(\theta)\bigg\}\leq e^{\rm I}\leq e^{\rm lim},\label{P4C}.			
 \end{eqnarray}
We show these results in the following two lemmas.
\begin{mydef1}\label{lemma3}
For a given $e^{\rm I}>0$, the optimal solution for optimization problem (P3) is $\theta=\max\{\theta^*,\theta^0\}$, where $\theta^*$ satisfies
\begin{eqnarray}
\eta e^{\rm I}+\mathcal{E}_{\rm D}(\theta^*)-(\theta^*-1)\theta^*\frac{\partial \mathcal{E}_{\rm D}(\theta)}{\partial \theta}\Big|_{\theta=\theta^*}=0,\label{theta_star}
\end{eqnarray}
and $\theta^0$ satisfies
\begin{eqnarray}
\mathcal{E}_{\rm D}(\theta^0)=\max\bigg\{0,\frac{\eta e^{\rm avg}-g}{e^{\rm lim}-e^{\rm avg}} e^{\rm lim}-\frac{\eta e^{\rm lim}-g}{e^{\rm lim}-e^{\rm avg}} e^{\rm I}\bigg\}.
\end{eqnarray}
\end{mydef1}

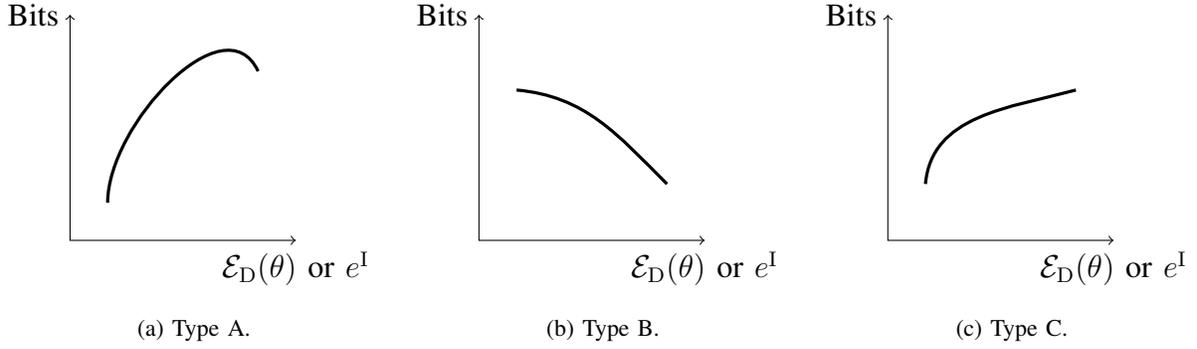
\begin{figure*}[!h]
	\centering
	\subfloat[Type A.]{\centering
			\begin{tikzpicture}[xscale = 0.5, yscale = 0.5]
			\draw[<->] (6,0) node[below]{$\mathcal{E}_{\rm D}(\theta)$ or $e^{\rm I}$} -- (0,0) --
			(0,6) node[left]{Bits};
\			\draw[very thick] (1,1) to [out=90,in=115] (5,4.5);
			\end{tikzpicture}
		}
	\hfil
	\subfloat[Type B.]{\centering
			\begin{tikzpicture}[xscale = 0.5, yscale = 0.5]
			\draw[<->] (6,0) node[below]{$\mathcal{E}_{\rm D}(\theta)$ or $e^{\rm I}$} -- (0,0) --
			(0,6) node[left]{Bits};
			\draw[very thick] (1,4) to [out=-5,in=135] (5,1.5);
			\end{tikzpicture}
		}
	\hfil
	\subfloat[Type C.]{\centering
		\begin{tikzpicture}[xscale = 0.5, yscale = 0.5]
		\draw[<->] (6,0) node[below]{$\mathcal{E}_{\rm D}(\theta)$ or $e^{\rm I}$} -- (0,0) --
		(0,6) node[left]{Bits};
		\draw[very thick] (1,1.5) to [out=85,in=195] (5,4);
		\end{tikzpicture}
		}	
	\caption{Illustrations of curves for different types.}
	\label{fig1}

\end{figure*}

\begin{proof}
For convenience of expression, we set $C_{e^{\rm I}}=(\eta e^{\rm avg}-g)C(e^{\rm I})$, which is a constant in this problem and does not affect optimal $\theta$. Taking derivative of objective function in terms of $\theta$, we can derive
\begin{eqnarray}
\frac{\partial \Big( C_{e^{\rm I}}\frac{\theta-1}{\theta(\eta e^{\rm I}+\mathcal{E}_{\rm D}(\theta))}\Big)}{\partial \theta}=C_{e^{\rm I}}\frac{\mathcal{M}(\theta)}{(\theta(\eta e^{\rm I}+\mathcal{E}_{\rm D}(\theta)))^2},
\end{eqnarray}
where
\begin{eqnarray}
\mathcal{M}(\theta)=\eta e^{\rm I}+\mathcal{E}_{\rm D}(\theta)-(\theta-1)\theta\frac{\partial \mathcal{E}_{\rm D}(\theta)}{\partial \theta}.
\end{eqnarray}
Firstly, we can obtain that
\begin{eqnarray}
\frac{\partial \mathcal{M}(\theta)}{\partial \theta}=-(2\theta-2)\frac{\partial \mathcal{E}_{\rm D}(\theta)}{\partial \theta}-(\theta^2-\theta)\frac{\partial^2 \mathcal{E}_{\rm D}(\theta)}{\partial \theta^2},
\end{eqnarray}
which is always non-positive when $\theta \geq 1$, so $\mathcal{M}(\theta)$ is non-increasing.
Then, since $\mathcal{E}_{\rm D}(\theta)$ is non-decreasing convex function, we must have $\frac{\partial \mathcal{E}_{\rm D}(\theta)}{\partial \theta}$ is non-decreasing, so we have
\begin{eqnarray}
\mathcal{E}_{\rm D}(\theta) =\int_{1}^{\theta}\frac{\partial \mathcal{E}_{\rm D}(\theta)}{\partial \theta}d\theta\leq (\theta-1)\frac{\partial \mathcal{E}_{\rm D}(\theta)}{\partial \theta},
\end{eqnarray}
and
\begin{eqnarray}
\mathcal{M}(\theta)\leq \eta e^{\rm I}-(\theta-1)^2\frac{\partial \mathcal{E}_{\rm D}(\theta)}{\partial \theta}.\label{Mtheta}
\end{eqnarray}
From \eqref{Mtheta} we can see, $\mathcal{M}(\theta)<0$ as $\theta\rightarrow +\infty$. Since we also have $\mathcal{M}(1)=\eta e^{\rm I}>0$, there must exist an optimal $\theta^*$ which satisfies \eqref{theta_star} and maximizes \eqref{P3O}. If we do not consider the inequality \eqref{P3C}, $\theta^*$ should be optimal solution for (P3). However, $\theta^*$ may not satisfy this inequality constraint. Thus, we can see (P3) is maximized at $\max\{\theta^*, \theta^0\}$.
\end{proof}
\begin{mydef5} \label{corollary1}
	When $e^{\rm I}\geq e^{\rm lim}(\eta e^{\rm avg}-g)/(\eta e^{\rm lim}-g)$, (P3) is maximized at $\theta^*$.
\end{mydef5}
\begin{proof}
	It is easy to see that in this case, $\theta^0=1$. However, since $\mathcal{M}(1)>0$, we must have $\theta^*>1$.
\end{proof}
\begin{mydef1}\label{lemma4}
For a given $\theta>1$, the optimal solution for optimization problem (P4) is $e^{\rm I} = \min\big\{ \max \big\{e^{\rm I*},e^{\rm I0}\big\},e^{\rm lim}\big\}$, where	$e^{\rm I*}$ satisfies
	\begin{eqnarray}
	\frac{\partial C(e^{\rm I})}{\partial e^{\rm I}}\bigg|_{e^{\rm I}=e^{\rm I*}}\cdot(\eta e^{\rm I*}+\mathcal{E}_{\rm D}(\theta))-\eta C(e^{\rm I*})=0,\label{ei_star}
	\end{eqnarray}
	and $e^{\rm I0}$ satisfies
	\begin{eqnarray}
	e^{\rm I0}=\max\bigg\{0,\frac{\eta e^{\rm avg}-g}{\eta e^{\rm lim}-g}e^{\rm lim}-\frac{e^{\rm lim}-e^{\rm avg}}{\eta e^{\rm lim}-g}\mathcal{E}_{\rm D}(\theta)\bigg\}.
	\end{eqnarray}
\end{mydef1}
\begin{proof}
For convenience of expression, we set $C_{\theta}= \frac{\theta-1}{\theta}(\eta e^{\rm avg}-g)$, which is a constant in this problem and does not affect optimal $e^{\rm I}$. Taking derivative of objective function in terms of $e^{\rm I}$, we can derive
	\begin{eqnarray}
	\frac{\partial \Big( C_{\theta}\frac{C(e^{\rm I})}{\eta e^{\rm I}+\mathcal{E}_{\rm D}(\theta)}\Big)}{\partial e^{\rm I}}=C_{\theta}\frac{\mathcal{N}(e^{\rm I})}{(\eta e^{\rm I}+\mathcal{E}_{\rm D}(\theta))^2},
	\end{eqnarray}
	where
	\begin{eqnarray}
	\mathcal{N}(e^{\rm I})=\frac{\partial C(e^{\rm I})}{\partial e^{\rm I}}\cdot(\eta e^{\rm I}+\mathcal{E}_{\rm D}(\theta))-\eta C(e^{\rm I}).
	\end{eqnarray}
	We can see that
	\begin{eqnarray}
	\frac{\partial \mathcal{N}(e^{\rm I})}{\partial e^{\rm I}}=\frac{\partial^2 C(e^{\rm I})}{\partial e^{\rm I2}}\cdot(\eta e^{\rm I}+\mathcal{E}_{\rm D}(\theta)).
	\end{eqnarray}
	We can prove that $\frac{\partial^2 C(e^{\rm I})}{\partial e^{\rm I2}}\leq 0$, please refer to Appendix A. Hence, $\mathcal{N}(e^{\rm I})$ in non-increasing. In addition, we can derive that
	\begin{eqnarray}
	\frac{\partial C(e^{\rm I})}{\partial e^{\rm I}}=\Big[\log_2\big(1-Q\big(\sqrt{2e^{\rm I}}\big)\big)-\log_2\big(Q\big(\sqrt{2e^{\rm I}}\big)\big)\Big]\cdot\frac{1}{\sqrt{4\pi}}e^{-e^{\rm I}}(e^{I})^{-0.5}.
	\end{eqnarray} Since $\frac{\partial C(e^{\rm I})}{\partial e^{\rm I}}|_{e^{\rm I}=0}>0$ and $C(0)=0$, we can get $\mathcal{N}(0)>0$. In addition, $\frac{\partial C(e^{\rm I})}{\partial e^{\rm I}}\cdot(\eta e^{\rm I}+\mathcal{E}_{\rm D}(\theta))\rightarrow 0$ and $-\eta C(e^{\rm I})\rightarrow -\eta$ as $e^{\rm I}\rightarrow + \infty$, so we can see that there must exist an optimal $e^{\rm I*}$ which satisfies \eqref{ei_star}. Similarly, if we do not consider the inequality constraints in (P4), $e^{\rm I*}$ should be optimal. When we consider the inequality constraints, we can see (P4) is maximized at $\min\big\{ \max \big\{e^{\rm I*},e^{\rm I0}\big\},e^{\rm lim}\big\}$.
\end{proof}
\begin{mydef5} \label{corollary2}
	When $\mathcal{E}_{\rm D}(\theta) \geq e^{\rm lim}(\eta e^{\rm avg}-g)/(e^{\rm lim}-e^{\rm avg})$, (P4) is maximized at $\min \{e^{\rm I*},e^{\rm lim}\}$.
\end{mydef5}
\begin{proof}
	It is easy to see that because in this case, $e^{\rm I0}=0$. However, since $\mathcal{N}(0)>0$, we must have $e^{\rm I*}>0$.
\end{proof}
In Fig. \ref{fig1}, we plot the values of \eqref{P3O} for different values of $\mathcal{E}_{\rm D}(\theta)$, and the values of \eqref{P4O} for different values of $e^{\rm I}$. Based on the analysis on Lemma \ref{lemma3} and Lemma \ref{lemma4}, these curves can be divided into three types, we call Type A, Type B, and Type C, respectively. In Type A, the curve firstly increases, then decreases, which has a single mode. In Type B, the curve is monotone non-increasing, and in Type C, the curve is monotone non-decreasing.

In Fig. \ref{fig2}, we draw the region of ($e^{\rm I}$, $\mathcal{E}_{\rm D}(\theta)$) under constraints \eqref{ineq_constrain_1}, \eqref{ineq_constrain_2}, and \eqref{ineq_constrain_4}. The coordinates of point A are ($e^{\rm lim}(\eta e^{\rm avg}-g)/(\eta e^{\rm lim}-g)$, 0) and the coordinates of point B are (0, $e^{\rm lim}(\eta e^{\rm avg}-g)/(e^{\rm lim}-e^{\rm avg})$). When we `observe' the value of \eqref{P2O} from a vertical or horizontal line, it actually corresponds to one of curves introduced above. We can have the following observations:
\begin{figure*}[!h]
\centering
\begin{tikzpicture}[xscale = 1, yscale = 1]
\draw[-][thick] (5,6) -- (5,0) node[below] {$e^{\rm lim}$};
\draw[fill] (2.5,0) circle [radius=0.025];
\node [below] at (2.5,0) {A};
\draw[fill] (0,3.5) circle [radius=0.025];
\node [left] at (0,3.5) {B};
\node [below] at (0,0) {0};
\draw[dashed] (0,0.5) -- (5,0.5);
\draw[dashed] (0,1) -- (5,1);
\draw[dashed] (0,1.5) -- (5,1.5);
\draw[ultra thick, dashed] (0,2) -- (5,2);
\node[right] at (5,2) {	
	\textcircled{4}};
\draw[dashed] (0,2.5) -- (5,2.5);
\draw[dashed] (0,3) -- (5,3);
\draw[dashed] (0,3.5) -- (5,3.5);
\draw[dashed] (0,4) -- (5,4);
\draw[ultra thick, dashed] (0,4.5) -- (5,4.5);
\node[right] at (5,4.5) {	
	\textcircled{3}};
\draw[dashed] (0,5) -- (5,5);
\draw[dashed] (0,5.5) -- (5,5.5);
\draw[dashed] (0.5,0) -- (0.5,6);
\draw[dashed] (1,0) -- (1,6);
\draw[dashed] (1.5,0) -- (1.5,6);
\draw[ultra thick, dashed] (2,0) -- (2,6);
\node[above] at (2,6) {	
	\textcircled{1}};
\draw[dashed] (2.5,0) -- (2.5,6);
\draw[ultra thick, dashed] (3,0) -- (3,6);
\node[above] at (3,6) {	
	\textcircled{2}};
\draw[dashed] (3.5,0) -- (3.5,6);
\draw[dashed] (4,0) -- (4,6);
\draw[dashed] (4.5,0) -- (4.5,6);
\draw [fill=white] (0,0) -- (0,3.5) -- (2.5,0);
\draw[thick,<->] (6,0) node[below]{$e^{\rm I}$} -- (0,0) --
(0,6) node[left]{$\mathcal{E}_{\rm D}(\theta)$};
\draw [thick] (0,3.5) -- (2.5,0);
\end{tikzpicture}
\caption{Region of ($e^{\rm I}$, $\mathcal{E}_{\rm D}(\theta)$) under constraints \eqref{ineq_constrain_1}, \eqref{ineq_constrain_2}, and \eqref{ineq_constrain_4}.}	
\label{fig2} 
\end{figure*}
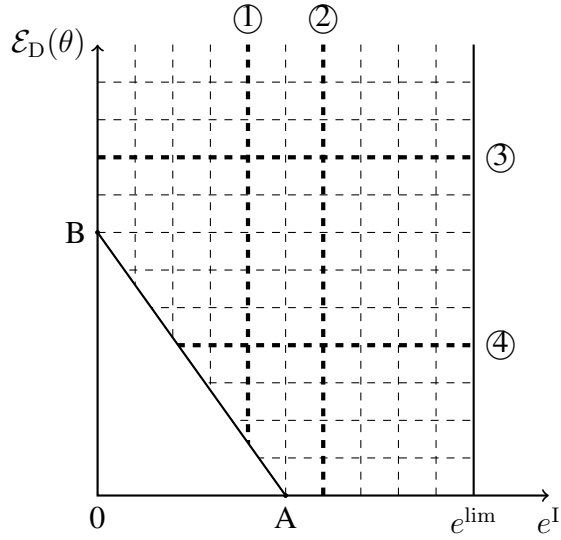
\begin{enumerate}
	\item When we `observe' from vertical line between (0, 0) and point A, like line \textcircled{1}, according to Lemma \ref{lemma3}, the curve can be either Type A or Type B.
	\item When we `observe' from vertical line between point A and ($e^{\rm lim}$, 0), like line \textcircled{2}, the curve can only be Type A. It is due to the fact that, in this case, $e^{\rm I}\geq e^{\rm lim}(\eta e^{\rm avg}-g)/(\eta e^{\rm lim}-g)$, so in Lemma \ref{lemma3}, \eqref{P3C} becomes $\mathcal{E}_{\rm D}(\theta)\geq 0$. However, since $\mathcal{M}(1)>0$, the curve must increase for a while, making Type B impossible.
	\item When we `observe' from horizontal line above point B, like line \textcircled{3}, the curve can be either Type A or Type C. Similarly, in this case, $\mathcal{E}_{\rm D}(\theta)\geq e^{\rm lim}(\eta e^{\rm avg}-g)/(e^{\rm lim}-e^{\rm avg})$, so in Lemma \ref{lemma4}, \eqref{P4C} becomes $0\leq e^{\rm I}\leq e^{\rm lim}$. Since $\mathcal{N}(0)>0$, the curve must increase for a while, making Type B impossible.
	\item When we `observe' from horizontal line between (0, 0) and point B, like line \textcircled{4}, according to Lemma \ref{lemma4}, the curve can be any type.
\end{enumerate}
Obviously, if a point in the region of ($e^{\rm I}$, $\mathcal{E}_{\rm D}(\theta)$) is optimal, when we `observe' from vertical line that passes through this point, it should maximize \eqref{P3O} for the given $e^{\rm I}$. Similarly, we `observe' from horizontal line that passes through this point, it should maximize \eqref{P4O} for the given $\mathcal{E}_{\rm D}(\theta)$. Then, we analyze on the possible places that the optimal point lies in and the conditions the optimal point should satisfy.
\begin{figure*}[t]
	\centering
	\begin{tikzpicture}[xscale = 1, yscale = 1]
	\draw[-][ultra thick,green] (5,6) -- (5,0);
	\node[below] at (5,0) {$e^{\rm lim}$};
	\draw[fill] (2.5,0) circle [radius=0.025];
	\node [below] at (2.5,0) {A};
	\draw[fill] (0,3.5) circle [radius=0.025];
	\node [left] at (0,3.5) {B};
	\node [below] at (0,0) {0};
	\draw[thick,<->] (6,0) node[below]{$e^{\rm I}$} -- (0,0) --
	(0,6) node[left]{$\mathcal{E}_{\rm D}(\theta)$};
	\draw [ultra thick, blue] (0,3.5) -- (2.5,0);
	\draw [fill=red,red] (4.97,5.6) -- (4.97,0.03) -- (2.53,0.03) -- (0.03,3.5) -- (0.03,5.6);
	\end{tikzpicture}
	\caption{Possible optimal point in the region.}	
	\label{fig3} 
\end{figure*}
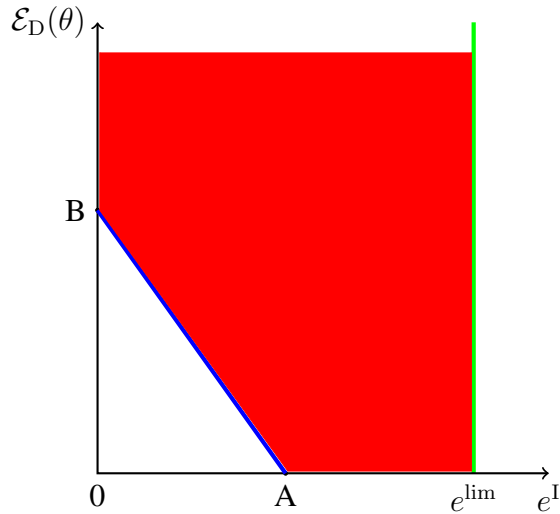

We use Fig. \ref{fig3} to illustrate.  Since $\mathcal{M}(1)>0$ in Lemma \ref{lemma3} and $\mathcal{N}(0)>0$ in Lemma \ref{lemma4}, the optimal point cannot lie in the segment between point A and ($e^{\rm lim}$, 0), and the ray that starts from point B along Y axis. Then we divide possible area into three cases.
\begin{enumerate}
	\item The optimal point is inside the red area.
	\item The optimal point is on the green ray that starts from point ($e^{\rm lim}$, 0) along Y axis. 
	\item The optimal point is on the blue segment between point A and point B.
\end{enumerate}
In case 1), since the optimal point in not on the boundary, when we `observe' from either vertical or horizontal line which passes through this point, the curve should be Type A. In case 2), the optimal point is the point achieve maximum when we `observe' from the green ray. In case 3), we must have
\begin{eqnarray}
\mathcal{E}_{\rm D}(\theta) = \frac{\eta e^{\rm avg}-g}{e^{\rm lim}-e^{\rm avg}} e^{\rm lim}-\frac{\eta e^{\rm lim}-g}{e^{\rm lim}-e^{\rm avg}} e^{\rm I}.
\end{eqnarray}
Based on these observations, we provide the following theorem to solve optimization problem (P2).
\begin{mydef2} \label{theorem2}
	The necessary condition that ($e^{\rm I}$, $\theta$) is optimal solution for (P2) is that it should satisfy at least one of the following sets of conditions:
\begin{enumerate}
	\item[(a)] 
	\begin{eqnarray}\label{Casea}
		\left\{\begin{array}{l} \eta e^{\rm I}+\mathcal{E}_{\rm D}(\theta)-(\theta-1)\theta\frac{\partial \mathcal{E}_{\rm D}(\theta)}{\partial \theta}=0,\\ 
		\frac{\partial C(e^{\rm I})}{\partial e^{\rm I}}\cdot(\eta e^{\rm I}+\mathcal{E}_{\rm D}(\theta))-\eta C(e^{\rm I})=0.\end{array} \right.
	\end{eqnarray}
	\item[(b)] 
	\begin{eqnarray}\label{Caseb}
		\left\{\begin{array}{l} 	\eta e^{\rm I}+\mathcal{E}_{\rm D}(\theta)-(\theta-1)\theta\frac{\partial \mathcal{E}_{\rm D}(\theta)}{\partial \theta}=0,\\	
		e^{\rm I}-e^{\rm lim}=0.
		\end{array} \right.
	\end{eqnarray}
	\item[(c)] 
	\begin{eqnarray}\label{Casec}
		\left\{\begin{array}{l}     \mathcal{E}_{\rm D}(\theta)+\frac{\eta e^{\rm lim}-g}{e^{\rm lim}-e^{\rm avg}} e^{\rm I} - \frac{\eta e^{\rm avg}-g}{e^{\rm lim}-e^{\rm avg}} e^{\rm lim}= 0,\\
		(e^{\rm lim}-e^{\rm I})C(e^{\rm I})+(\theta^2-\theta)(e^{\rm lim}-e^{\rm I})\frac{\partial \tilde{C}(\theta)}{\partial \theta}=(\theta^2-\theta)C(e^{\rm I})\frac{e^{\rm lim}-e^{\rm avg}}{\eta e^{\rm lim}-g}\frac{\partial \mathcal{E}_{\rm D}(\theta)}{\partial \theta},
		\end{array} \right.
	\end{eqnarray}
	where $\tilde{C}(\theta)=C(\tilde{e}^{\rm I}(\theta))$  and $\tilde{e}^{\rm I}(\theta) = \frac{\eta e^{\rm avg}-g}{\eta e^{\rm lim}-g}e^{\rm lim}-\frac{e^{\rm lim}-e^{\rm avg}}{\eta e^{\rm lim}-g}\mathcal{E}_{\rm D}(\theta)$, which are channel capacity and energy per symbol expressed in terms of $\theta$, respectively.
\end{enumerate}
\end{mydef2}
\begin{proof}
Assume ($\underline{e}^{\rm I}$, $\underline{\theta}$) is the optimal solution, according to Lemma \ref{lemma3} and Corollary \ref{corollary1}, for the first equation of \eqref{Casea} and the first equation of \eqref{Casec}, at least one must hold.
\begin{enumerate}
	\item If the first equation of \eqref{Casea} holds, but the first equation of \eqref{Casec} does not hold, according to Lemma \ref{lemma4} and Corollary \ref{corollary2}, we must have that, for the second equation of \eqref{Casea} and the second equation of \eqref{Caseb}, at least one must hold. It means that at least one of \eqref{Casea} and \eqref{Caseb} must hold.
	\item If the first equation of \eqref{Casec} holds, by expressing $e^{\rm I}$ in terms of $\theta$, (P2) becomes
\\* $\rm (P5)$
\begin{eqnarray}
			\max\limits_{\theta}&\quad&\frac{\theta-1}{\theta}\cdot \frac{e^{\rm lim}-e^{\rm avg}}{e^{\rm lim}-\tilde{e}^{\rm I}(\theta)}\cdot\tilde{C}(\theta),\\{\rm s.t.}&\quad& 0\leq \mathcal{E}_{\rm D}(\theta)\leq \frac{\eta e^{\rm avg}-g}{e^{\rm lim}-e^{\rm avg}}\cdot e^{\rm lim},
\end{eqnarray}
where the definitions of $\tilde{C}(\theta)$ and $\tilde{e}^{\rm I}(\theta)$ are given above. To find the optimized $\theta$, we can obtain that 
\begin{eqnarray}
\frac{\partial \tilde{e}^{\rm I}(\theta)}{\partial \theta}&=&-\frac{e^{\rm lim}-e^{\rm avg}}{\eta e^{\rm lim}-g}\cdot \frac{\partial \mathcal{E_{\rm D}(\theta)}}{\partial \theta}\leq 0,\\
\frac{\partial^2 \tilde{e}^{\rm I}(\theta)}{\partial \theta^2}&=&-\frac{e^{\rm lim}-e^{\rm avg}}{\eta e^{\rm lim}-g}\cdot \frac{\partial^2 \mathcal{E_{\rm D}(\theta)}}{\partial \theta^2}\leq 0,
\end{eqnarray}
and
\begin{eqnarray}
\frac{\partial \tilde{C}(\theta)}{\partial \theta}&=&\frac{\partial C(\tilde{e}^{\rm I}(\theta))}{\partial \tilde{e}^{\rm I}(\theta)}\cdot\frac{\partial \tilde{e}^{\rm I}(\theta)}{\partial \theta}\leq0,\\
\frac{\partial^2 \tilde{C}(\theta)}{\partial \theta^2}&=&\frac{\partial^2 C(\tilde{e}^{\rm I}(\theta))}{\partial \tilde{e}^{\rm I}(\theta)^2}\cdot\bigg(\frac{\partial \tilde{e}^{\rm I}(\theta)}{\partial \theta}\bigg)^2+\frac{\partial C(\tilde{e}^{\rm I}(\theta))}{\partial \tilde{e}^{\rm I}(\theta)}\cdot\frac{\partial^2 \tilde{e}^{\rm I}(\theta)}{\partial \theta^2}\leq0，
\end{eqnarray}
Define $f(\theta)=\frac{(\theta-1)\tilde{C}(\theta)}{\theta(e^{\rm lim}-\tilde{e}^{\rm I}(\theta))}$, so the objective function of (P5) can be written as $(e^{\rm lim}-e^{\rm avg})\cdot f(\theta)$. Taking derivative of $f(\theta)$ in terms of $\theta$, we can obtain
\begin{eqnarray}
\frac{\partial f(\theta)}{\partial \theta}=\frac{h(\theta)}{\theta^2(e^{\rm lim}-\tilde{e}^{\rm I}(\theta))^2},
\end{eqnarray}
where 
\begin{eqnarray}
h(\theta)=(e^{\rm lim}-\tilde{e}^{\rm I}(\theta))\tilde{C}(\theta)+(\theta^2-\theta)(e^{\rm lim}-\tilde{e}^{\rm I}(\theta))\frac{\partial \tilde{C}(\theta)}{\partial \theta}+(\theta^2-\theta)\tilde{C}(\theta)\frac{\partial \tilde{e}^{\rm I}(\theta)}{\partial \theta}.
\end{eqnarray}
Since
\begin{eqnarray}
\frac{\partial h(\theta)}{\partial \theta}&=&2\theta(e^{\rm lim}-\tilde{e}^{\rm I}(\theta))\frac{\partial \tilde{C}(\theta)}{\partial \theta}+(\theta^2-\theta)(e^{\rm lim}-\tilde{e}^{\rm I}(\theta))\frac{\partial^2 \tilde{C}(\theta)}{\partial \theta^2}\nonumber\\&+&(2\theta-2)\tilde{C}(\theta)\frac{\partial \tilde{e}^{\rm I}(\theta)}{\partial \theta}+(\theta^2-\theta)\tilde{C}(\theta)\frac{\partial^2 \tilde{e}^{\rm I}(\theta)}{\partial \theta^2}\leq 0,\label{Q54}
\end{eqnarray}
and
\begin{eqnarray}
h(1)&=& \frac{\eta e^{\rm lim}-\eta e^{\rm avg}}{\eta e^{\rm lim}-g}e^{\rm lim}\cdot C\bigg(\frac{\eta e^{\rm avg}-g}{\eta e^{\rm lim}-g}e^{\rm lim}\bigg)> 0,\label{Q55}\\
h(\theta')&=& (\theta'^2-\theta')e^{\rm lim}\cdot\frac{\partial C(e^{\rm I})}{\partial e^{\rm I}}\bigg|_{e^{\rm I}=0}\cdot\bigg(-\frac{e^{\rm lim}-e^{\rm avg}}{\eta e^{\rm lim}-g}\bigg)\cdot \frac{\partial \mathcal{E_{\rm D}(\theta)}}{\partial \theta}\bigg|_{\theta = \theta'} < 0\label{Q56},
\end{eqnarray}
where $\mathcal{E}_{\rm D}(\theta')=\frac{\eta e^{\rm avg}-g}{e^{\rm lim}-e^{\rm avg}}\cdot e^{\rm lim}$. From \eqref{Q54}-\eqref{Q56} we can see that the optimal solution $\underline{\theta}$ must satisfy 
\begin{eqnarray}
h(\underline{\theta})=0.\label{case_3_eqn2}
\end{eqnarray}
Thus, \eqref{Casec} must be satisfied.
\end{enumerate}
This completes the proof of the theorem.
\end{proof}
\begin{mydef3}\label{remark1}
	When \eqref{Casec} holds, due to \eqref{expressing_ee}, we can obtain 
	\begin{eqnarray}
	e^{\rm E}=e^{\rm lim},
	\end{eqnarray}
	which means that the transmitting power for energy harvesting is maximized.
\end{mydef3}

Thus, we can respond to the question asked in Section I.
The three equation groups can be explained as
\begin{enumerate}
	\item In case (b), we have $e^{\rm I}=e^{\rm lim}$, which means that we should maximize the transmitting power for information transmission.
	\item In case (c), due to Remark \ref{remark1}, we have $e^{\rm E}=e^{\rm lim}$, which means that we should maximize the transmitting power for energy harvesting.
	\item In case (a), we maximize neither transmitting power for information transmission nor transmitting power for energy harvesting, but make a trade-off between them.
\end{enumerate} 
Due to case (b) and case (c), we surprisingly find that the ideas of maximizing power for energy harvesting and information transmission both have the potential to be optimal. In addition, for some cases we should make a trade-off between them, which yields case (a). One decisive factor that determines which one is optimal is the form of $\mathcal{E}_{\rm D}(\theta)$, which inherently reflects the decoding scheme used and hardware implementation. 

As mentioned in Section I, in practical systems, $\mathcal{E}_{\rm D}(\theta)$ can be obtained by experiments or simulations, so it is possible to do an off-line optimization based on Theorem \ref{theorem2} for different values of $g$ beforehand. Hence, the devices can obtain the best scheme by simply evaluating $g$ when they are working. The global optimal solution can be obtained by solving \eqref{Casea}-\eqref{Casec} and selecting the one maximizing \eqref{P2O} from all possible solutions. Notice that we convert the problem from solving non-convex optimization problem into solving equation groups, and for many specific forms of $\mathcal{E}_{\rm D}(\theta)$, some simple-form or even closed-form solutions can be obtained. Hence, the complexity may decrease. We denote the value of \eqref{P2O} as $\mathcal{O}(\theta,{e}^{\rm I})$. The algorithm to solve (P2) can be summarized as Algorithm \ref{al1}.

\begin{algorithm}[t]
	\caption{Finding optimal solution for (P2)}
	\label{al1}
	\begin{algorithmic}
		\State{Solve equations \eqref{Casea}.}
		\State{Solve equations \eqref{Caseb}.}
		\State{Solve equations \eqref{Casec}.}
		\State{Set $\theta^{\rm opt}=0,{e}^{\rm Iopt}=0, fval =0$}
		\For{all solutions $({e}^{\rm I}, \theta)$ above}
		\If{$({e}^{\rm I}, \theta)$ satisfies \eqref{ineq_constrain_1}, \eqref{ineq_constrain_2}, \eqref{ineq_constrain_4} {\bf and} $fval < \mathcal{O}(\theta, {e}^{\rm I})$}
		\State{Set $\theta^{\rm opt}=\theta$,${e}^{\rm Iopt}={e}^{\rm I}$, $fval = \mathcal{O}(\theta^{\rm opt},{e}^{\rm Iopt})$.}
		\EndIf
		\EndFor
	\end{algorithmic}
\end{algorithm}

\section{Transmission over Multiple Blocks}
In this section,  we extend our results to multiple transmission blocks. We assume there are $N$ blocks in total and use the subscript $i$ to denote parameters for the $i$th block. In addition, the average and peak power constraints are the same for all blocks and we have $\eta e^{\rm avg}-g_i\geq 0$, $i = 1,\dots,N$. To maximize the total number of bits decoded over $N$ blocks, the optimization problem becomes
 \\* $\rm (P6)$
\begin{eqnarray}
\max\limits_{\bm{\alpha}, \bm{R}, \bm{e^{\rm E}}, \bm{e^{\rm I}}}&\quad&\sum_{i=1}^{N}(1-\alpha_i)R_i,\\{\rm s.t.}&\quad&\sum_{i=1}^{k}(1-\alpha_i)\mathcal{E}_{\rm D}(\theta_i)+g_i\leq \eta\alpha_i e_i^{\rm E},\quad \forall k,
\label{P6_first_contraints}\\{}&\quad& \alpha_i e_i^{\rm E} + (1-\alpha_i) e_i^{\rm I} \leq e^{\rm avg}, \quad \forall i,\\{}&\quad& \theta_i = \frac{C(e^{\rm I}_i)}{C(e^{\rm I}_i)-R_i}	\quad \forall i,
\\{}&\quad& 0\leq e_i^{\rm E}\leq e^{\rm lim}, \quad \forall i,
\\{}&\quad& 0\leq e_i^{\rm I}\leq e^{\rm lim}, \quad \forall i,
\\{}&\quad& 0\leq \alpha_i\leq 1, \quad \forall i,
\\{}&\quad& 0\leq R_i\leq C(e_i^{\rm I}),  \quad \forall i,
\end{eqnarray}
where $\bm{\alpha}=\{\alpha_1,\dots,\alpha_N\}$, $\bm{R}=\{R_1,\dots,R_N\}$, $\bm{e^{\rm E}}=\{e^{\rm E}_1,\dots,e^{\rm E}_N\}$, and $\bm{e^{\rm I}}=\{e^{\rm I}_1,\dots,e^{\rm I}_N\}$. Notice that \eqref{P6_first_contraints} becomes a group of energy-causality constraints, since the energy harvested can be stored for use in the following blocks. To decompose these constraints, we introduce $Ti$ that $(1-\alpha_i)\mathcal{E}_{\rm D}\Big(\frac{C_i}{C_i-R_i}\Big)+g_i+T_i = \eta\alpha_i e_i^{\rm E}$. Similar to Section \ref{singleblock}, we can convert (P6) into an equivalent optimization problem (P7), which is given as
 \\* $\rm (P7)$
\begin{eqnarray}
\max\limits_{\bm{\theta}, \bm{e^{\rm I}}, \bm{T}}&\quad&\sum_{i=1}^{N}\frac{\theta_i-1}{\theta_i} (\eta e^{\rm avg}-g_i-T_i)\frac{C(e^{\rm I}_i)}{\eta e_i^{\rm I}+\mathcal{E}_{\rm D}(\theta_i)},\nonumber\\\\{\rm s.t.}&\quad& \sum_{i=1}^{k}T_i\geq 0, \quad \forall k, \label{mul_ineq_constrain_-1}\\&\quad& T_i \leq \eta e^{\rm avg} - g_i \label{mul_ineq_constrain_0} ,\quad \forall i, \\&\quad& 0\leq e_i^{\rm I}\leq e^{\rm lim},\quad \forall i,\label{mul_ineq_constrain_1}
\\{}&\quad& \theta_i \geq 1,\quad \forall i,\label{mul_ineq_constrain_2} 
\\{}&\quad& \mathcal{E}_{\rm D}(\theta_i)+\frac{\eta e^{\rm lim}-g_i-T_i}{e^{\rm lim}-e^{\rm avg}} e_i^{\rm I} \geq \frac{\eta e^{\rm avg}-g_i-T_i}{e^{\rm lim}-e^{\rm avg}} e^{\rm lim},\quad \forall i,\label{mul_ineq_constrain_4}			
\end{eqnarray}
where $\bm{\theta}=\{\theta_1,\dots,\theta_N\}$ and $\bm{T}=\{T_1,\dots, T_N\}$. Similar to \eqref{expressing_alpha}, to derive (P7) here we can obtain $\alpha_i = 1-\frac{\eta e^{\rm avg}-g_i-T_i}{\eta e_i^{\rm I}+\mathcal{E}_{\rm D}(\theta_i)}$, and \eqref{mul_ineq_constrain_0} is given due to the fact that $\alpha_i \leq 1$. Hence, we can see all the terms in objective function are non-negative.

\subsection{Upper Bound for (P7) and Conditions to Achieve the Bound}
(P7) is not easy to solve optimally. In this subsection, we give an upper bound for the solution to (P7). Then, we show in which scenarios this upper bound can be achieved. First, we define
\begin{eqnarray}
\tilde{\mathcal{O}}(\theta,e^{\rm I}) = \frac{\theta-1}{\theta}\cdot\frac{C(e^{\rm I})}{\eta e^{\rm I}+\mathcal{E}_{\rm D}(\theta)},
\end{eqnarray}
and introduce a new optimization problem
 \\* $\rm (P8)$
\begin{eqnarray}
\max\limits_{\theta, e^{\rm I}}&\quad&\tilde{\mathcal{O}}(\theta,e^{\rm I}),\\{\rm s.t.}&\quad& 0\leq e^{\rm I}\leq e^{\rm lim},\label{p8_constrain_1}
\\{}&\quad& \theta \geq 1.\label{p8_constrain_2} 			
\end{eqnarray}	
Based on the analysis in Theorem \ref{theorem2}, (P8) can be solved by Algorithm \ref{al1} without case (c) (i.e., solving equations \eqref{Casec}). We assume $\dot{\theta}$ and $\dot{e}^{\rm I}$ are optimal that maximize $\tilde{\mathcal{O}}(\theta,e^{I})$ under constraints \eqref{p8_constrain_1} and \eqref{p8_constrain_2}. We further define
\begin{eqnarray}
\dot{G}=\eta e^{\rm avg}-(\mathcal{E}_{\rm D}(\dot{\theta})+\eta \dot{e}^{\rm I})\cdot \frac{e^{\rm lim}-e^{\rm avg}}{e^{\rm lim}-\dot{e}^{\rm I}},
\end{eqnarray}  so it is easy to see $\mathcal{E}_{\rm D}(\dot{\theta})+\frac{\eta e^{\rm lim}-g-T}{e^{\rm lim}-e^{\rm avg}} \dot{e}^{\rm I} \geq \frac{\eta e^{\rm avg}-g-T}{e^{\rm lim}-e^{\rm avg}} e^{\rm lim}$ is equivalent to $T+g\geq \dot{G}$. Now we give the following lemma regarding the upper bound and the conditions to achieve the upper bound.
\begin{mydef1}\label{l1}
	An upper bound for the solution of (P7) is $\sum_{i=1}^{N}(\eta e^{\rm avg}-g_i)\tilde{\mathcal{O}}(\dot{\theta},\dot{e}^{\rm I})$. The upper bound can be achieved if and only if there exists a feasible set of ($\bm{\theta}, \bm{e^{\rm I}}, \bm{T}$) that satisfies $\theta_1=\cdots =\theta_{N}=\dot{\theta}$, $e^{\rm I}_1=\cdots =e_{N}^{\rm I}=\dot{e}^{\rm I}$, $\sum_{i=1}^{N}T_i=0$ and other constraints in (P7).
\end{mydef1}
\begin{proof}
	For any feasible solution for (P7), $\{\hat{\theta}_1,\dots,\hat{\theta}_N\}$, $\{\hat{e}_1^{\rm I},\dots,\hat{e}_N^{\rm I}\}$, and $\{\hat{T}_1,\dots, \hat{T}_N\}$, we can obtain
	\begin{eqnarray}
	&{}&\sum_{i=1}^{N}(\eta e^{\rm avg}-g_i-\hat{T}_i)\tilde{\mathcal{O}}(\hat{\theta}_i,\hat{e}^{\rm I}_i)\nonumber\\&\overset{(a)}{\leq}&\sum_{i=1}^{N}(\eta e^{\rm avg}-g_i-\hat{T}_i)\tilde{\mathcal{O}}(\dot{\theta},\dot{e}^{\rm I})\nonumber\\&=&\Bigg(\sum_{i=1}^{N}(\eta e^{\rm avg}-g_i)-\sum_{i=1}^{N}\hat{T}_i\Bigg)\tilde{\mathcal{O}}(\dot{\theta},\dot{e}^{\rm I})\nonumber\\&\overset{(b)}{\leq}& \sum_{i=1}^{N}(\eta e^{\rm avg}-g_i)\tilde{\mathcal{O}}(\dot{\theta},\dot{e}^{\rm I}),
	\end{eqnarray}
	where the equality in (a) holds if and only if $\theta_1=\cdots =\theta_{N}=\dot{\theta}$, $e^{\rm I}_1=\cdots =e_{N}^{\rm I}=\dot{e}^{\rm I}$ and the equality in (b) holds if and only if $\sum_{i=1}^{N}T_i=0$.
\end{proof}
To achieve the upper bound, there must exist a set of $\bm{T}$ can make $\sum_{i=1}^{N}T_i=0$, which is more strict than \eqref{mul_ineq_constrain_-1}, and since $\mathcal{E}_{\rm D}(\dot{\theta})+\frac{\eta e^{\rm lim}-g_i-T_i}{e^{\rm lim}-e^{\rm avg}} \dot{e}^{\rm I} \geq \frac{\eta e^{\rm avg}-g_i-T_i}{e^{\rm lim}-e^{\rm avg}} e^{\rm lim}$ holds for all $i$, we must have $T_i+g_i\geq \dot{G}$ for all $i$. Hence, whether there exists a solution satisfying the above constraints is highly dependent on the relationship between $\{g_1,\dots,g_N\}$ and $\dot{G}$.
\begin{mydef2} \label{theorem3}
The sufficient and necessary condition that the solution of (P7) is $\sum_{i=1}^{N}(\eta e^{\rm avg}-g_i)\tilde{\mathcal{O}}(\dot{\theta},\dot{e}^{\rm I})$ is
\begin{eqnarray} \label{theorem3_equality}
\sum_{i=k}^{N}g_i\geq (N-k+1)\dot{G},\quad \forall k. \label{theorem2_equality}
\end{eqnarray}
\end{mydef2}
\begin{proof}
	\subsubsection{Sufficiency}
	Firstly, we prove the sufficiency. Let 
	\begin{eqnarray}
	T_i=\left\{\begin{array}{ll} \max(0, \dot{G}-g_1),& i=1,\\ \max(-\sum_{j=1}^{i-1}T_j, \dot{G}-g_i),& i=2,\dots,N,\end{array} \right.
	\end{eqnarray}
	We can see $T_1\geq 0$ and for any $i\geq 2$, we have $\sum_{j=1}^{i}T_j=\sum_{j=1}^{i-1}T_j+T_i\geq\sum_{j=1}^{i-1}T_j-\sum_{j=1}^{i-1}T_j=0$, so \eqref{mul_ineq_constrain_-1} is satisfied. Then for any $i$, if $\dot{G}-g_i\leq0$, then $T_i\leq 0\leq \eta e^{\rm avg}-g_i$. Otherwise, $T_i=\dot{G}-g_i\leq\eta e^{\rm avg}-g_i$, so \eqref{mul_ineq_constrain_0} is also satisfied. And $T_i+g_i\geq \dot{G}-g_i+g_i=\dot{G}$, so $\theta_1=\cdots =\theta_{N}=\dot{\theta}$, $e^{\rm I}_1=\cdots =e_{N}^{\rm I}=\dot{e}^{\rm I}$ are feasible. To complete the proof of sufficiency, we only need to prove $\sum_{i=1}^{N}T_i=0$. Since $\sum_{i=1}^{N}T_i\geq 0$, our next work is to prove $\sum_{i=1}^{N}T_i \leq 0$.
	
	We assume there are totally $M$ blocks, indexed by $i_m$, $m=1,\dots,M$, in which $g_{i_m}\geq \dot{G}$ happens. It is easy to see $i_M=N$ because we can obtain $g_N\geq \dot{G}$ when we set $k=N$ in \eqref{theorem2_equality}. For $i'\neq i_m$, $m= 1,\dots, M$, $\dot{G}\geq g_{i'}$, so $T_{i'}=\dot{G}-g_{i'}$. For $T_{i_i},\dots,T_{i_M}$, we cannot decide their values, so we divide all possible combinations into $M+1$ cases illustrated by Fig. \ref{diving_all_possible_combinations}.
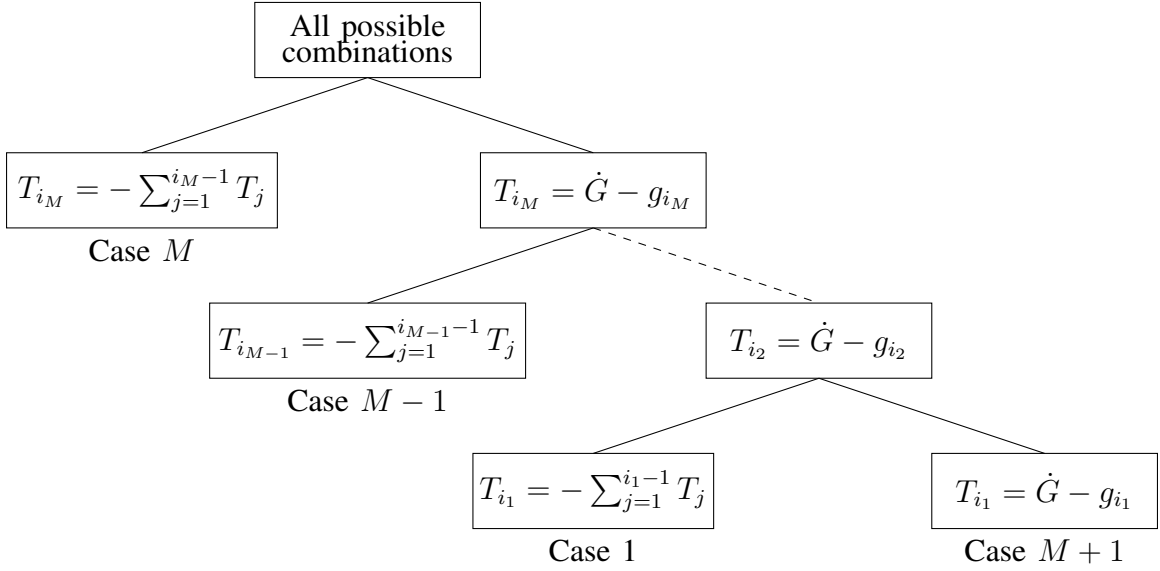
\begin{figure}
	\centering
	\begin{tikzpicture}

	\draw [black] (3,8) rectangle (6,7);
	\draw [black] (-0.3,6) rectangle (3.3,5);
	\draw [black] (6,6) rectangle (9,5);
	\draw [black] (2.4,4) rectangle (6.6,3);
	\draw [black] (9,4) rectangle (12,3);
	\draw [black] (5.9,2) rectangle (9.1,1);
	\draw [black] (12,2) rectangle (15,1);
	\draw [-] (4.5, 7) -- (1.5,6);
	\draw [-] (4.5, 7) -- (7.5,6);
	\draw [-] (7.5, 5) -- (4.5,4);
	\draw [dashed] (7.5, 5) -- (10.5,4);
	\draw [-] (10.5, 3) -- (7.5,2);
	\draw [-] (10.5, 3) -- (13.5,2);
	\node at (4.5,7.65) {All possible};
	\node at (4.5,7.35) {combinations};
	\node at (1.5,5.5) {$T_{i_M}=-\sum_{j=1}^{i_M-1}T_j$};
	\node [below] at (1.5,5) {Case $M$};
	\node at (7.5,5.5) {$T_{i_M}=\dot{G}-g_{i_M}$};
	\node at (4.5,3.5) {$T_{i_{M-1}}=-\sum_{j=1}^{i_{M-1}-1}T_j$};
	\node [below] at (4.5,3) {Case $M-1$};
	\node at (10.5,3.5) {$T_{i_{2}}=\dot{G}-g_{i_{2}}$};
	\node at (7.5,1.5) {$T_{i_{1}}=-\sum_{j=1}^{i_{1}-1}T_j$};
	\node [below] at (7.5,1) {Case 1};
	\node at (13.5,1.5) {$T_{i_{1}}=\dot{G}-g_{i_{1}}$};
	\node [below] at (13.5,1) {Case $M+1$};
	\end{tikzpicture}
	\caption{Dividing all possible combinations of values for $\{T_{i_1},\dots T_{i_M}\}$ into $M+1$ cases.}\label{diving_all_possible_combinations}
\end{figure}

For Case $m$, $m=1,\dots,M$, we have
\begin{eqnarray}
\sum_{i=1}^{N}T_i&=&\sum_{i=1}^{i_m-1}T_i+T_{i_m}+\sum_{i=i_m+1}^{N}T_i=\sum_{i=1}^{i_m-1}T_i-\sum_{i=1}^{i_m-1}T_i+\sum_{i=i_m+1}^{N}(\dot{G}-g_i)\nonumber\\&=&\sum_{i=i_m+1}^{N}(\dot{G}-g_i)\leq 0.
\end{eqnarray}
For Case $M+1$, we have
\begin{eqnarray}
\sum_{i=1}^{N}T_i= \sum_{i=1}^{N}(\dot{G}-g_i)\leq 0.
\end{eqnarray}
Thus, we show $\sum_{i=1}^{N}T_i\leq 0$ and we complete the proof of sufficiency.
\subsubsection{Necessity}
Then, we prove the necessity. From Lemma \ref{l1} we know, when $\sum_{i=1}^{N}(\eta e^{\rm avg}-g_i)\tilde{\mathcal{O}}(\dot{\theta},\dot{e}^{\rm I})$ can be achieved, we must have $\sum_{i=1}^{N}T_i=0$ and $T_i+g_i\geq \dot{G}$ for all $i$. Due to \eqref{mul_ineq_constrain_-1}, we must have
\begin{eqnarray}
\sum_{i=k}^{N}T_i = \sum_{i=1}^{N}T_i- \sum_{i=1}^{k-1}T_i=- \sum_{i=1}^{k-1}T_i\leq 0, \quad \forall k.
\end{eqnarray}
Then, since $g_i\geq \dot{G}-T_i$ for all $i$, we can obtain that
\begin{eqnarray}
\sum_{i=k}^{N}g_i\geq (N-k+1)\dot{G}-\sum_{i=k}^{N}T_i\geq (N-k+1)\dot{G}, \quad \forall k.
\end{eqnarray}
Thus, the necessity is proved.
\end{proof}
Intuitively, Theorem \ref{theorem3} says that, when $g_i$ is relatively large, if we decrease $g_i$ by one, the energy saved can be used to decode another $\tilde{\mathcal{O}}(\dot{\theta},\dot{e}^{\rm I})$ information bits. However, when $g_i$ becomes smaller, the additional number of bits decoded by decreasing $g_i$ may also be smaller due to the power constraints. In addition, we notice that  $\dot{G}$ can be non-positive, so \eqref{theorem3_equality} can alway hold for some cases.

\subsection{General Case}
For the general case, we can obtain a local optimal solution by optimizing ($\bm{e^{\rm I}}$, $\bm{\theta}$) and $\bm{T}$ iteratively.
\begin{enumerate}
	\item[I.] When $\bm{T}$ is fixed, (P7) can be decomposed into $N$ independent sub-optimization problems. It is easy to find that each sub-problem has the same form with (P2) and can be directly solved using Algorithm \ref{al1}.
	\item[II.] When $\bm{e^{\rm I}}$ and $\bm{\theta}$ are fixed. $\bm{T}$ can be solved by a standard linear programing (LP) method.
	\item[III.] Iteratively optimize ($\bm{e^{\rm I}}$, $\bm{\theta}$) and $\bm{T}$ until the conditions for convergence are satisfied.
\end{enumerate}

\section{Numerical Results}
In this section, we give numerical results to illustrate and verify the analysis presented in previous sections. We assume $\mathcal{E}_{\rm D}(\theta)=\theta \log_2\theta$, which coincides with the research on the decoding complexity of LDPC codes \cite{RLDPC2}.
\subsection{Single Block}
\begin{figure}[!h]
	\begin{center}
		\hspace*{-10mm}
		\includegraphics[scale=0.8]{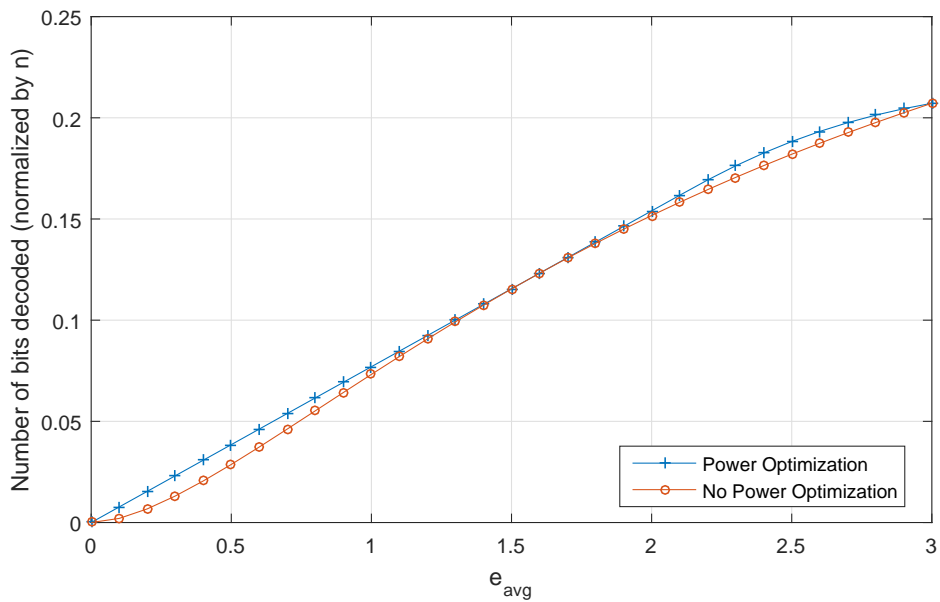}
	\end{center}
	\caption{Number of bits decoded with/without power optimization at the transmitter in a single block.}\label{single_block_constant_vs_optimized}
\end{figure}
\begin{figure}[!h]
	\begin{center}
		\hspace*{-10mm}
		\includegraphics[scale=0.8]{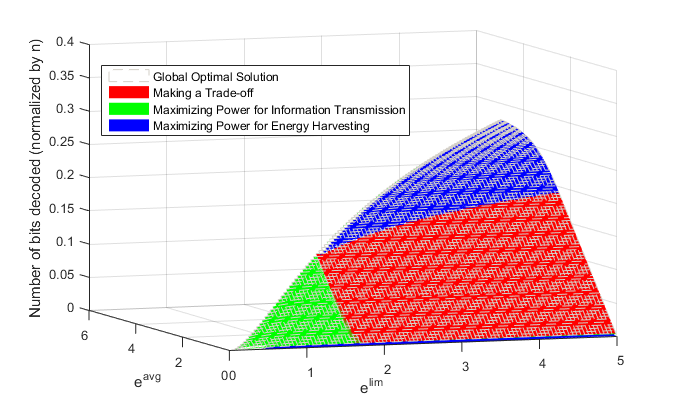}
	\end{center}
	\caption{Global optimal solution by numerically solving (P1) and solutions of case (a), case (b), and case (c) in Theorem \ref{theorem2}, under different values of $e^{\rm lim}$ and $e^{\rm avg}$.}\label{single_block_fig1_degree1eps}
\end{figure}
\begin{figure}[!h]
	\begin{center}
		\hspace*{-10mm}
		\includegraphics[scale=0.8]{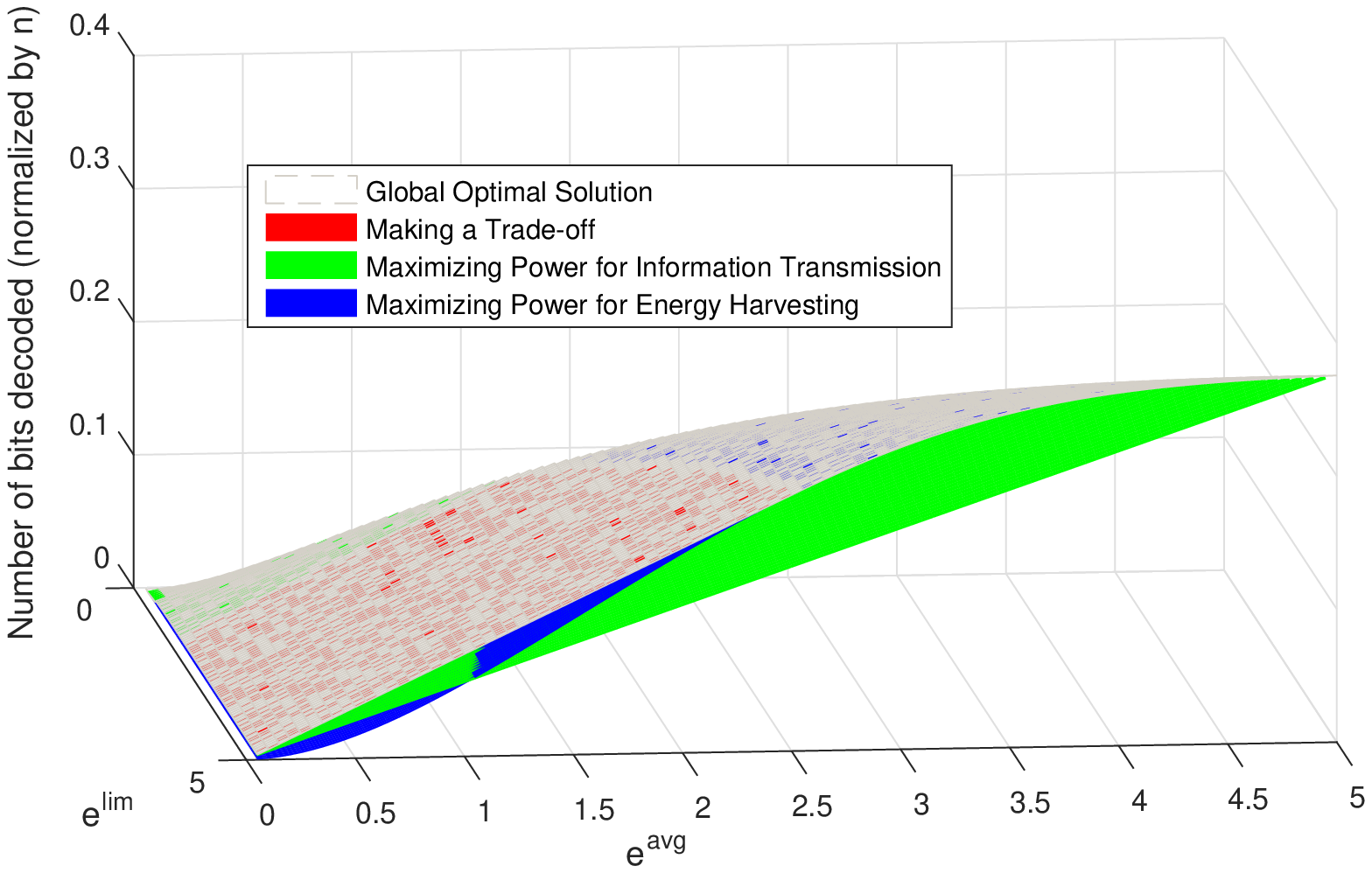}
	\end{center}
	\caption{Global optimal solution by numerically solving (P1) and solutions of case (a), case (b), and case (c) in Theorem \ref{theorem2}, under different values of $e^{\rm lim}$ and $e^{\rm avg}$.}\label{single_block_fig1_degree2eps}
\end{figure}
In this subsection, we consider transmission over a single block. Here, we assume $\eta = 0.5$ and $g=0$. Firstly, we show that considering power optimization at the transmitter can improve the performance. In, Fig. \ref{single_block_constant_vs_optimized}, we set $e^{\rm lim}=3$ and let $e^{\rm avg}$ changes from 0 to 3. When there is no power optimization, each symbol is transmitted at a predetermined constant power, irrespective of whether it is used for energy harvesting or information decoding, as assumed in our previous works \cite{RNi1,RNi2}. With power optimization (as studied in this paper), the powers used for energy harvesting and information transmission can be different. The numerical results in Fig. \ref{single_block_constant_vs_optimized} shows that the performance with power optimization is better than that without power optimization, e.g., at $e^{\rm avg}=0.5$, there is a 50\% increase in number of bits decoded.

\begin{figure}[!h]
	\begin{center}
		\hspace*{-10mm}
		\includegraphics[scale=0.8]{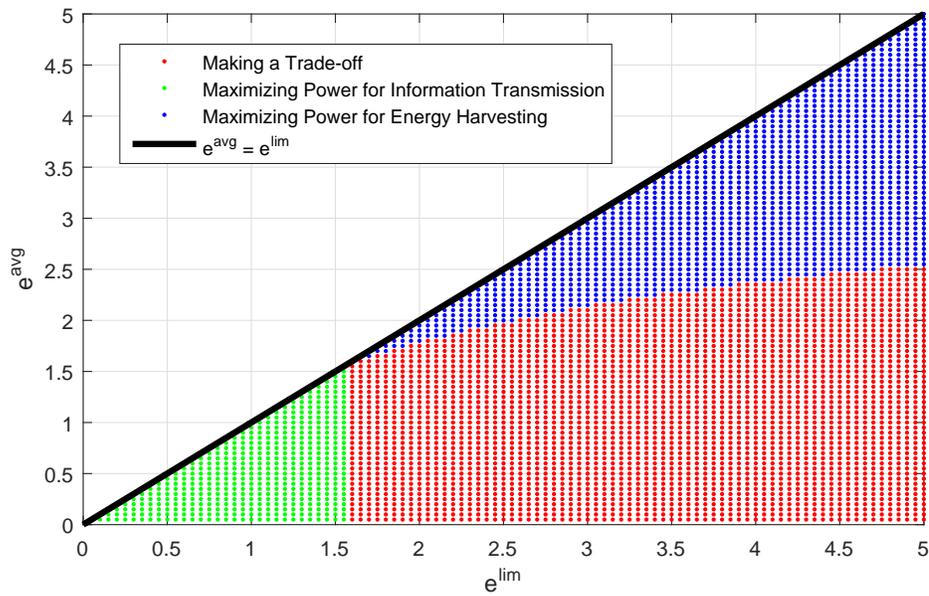}
	\end{center}
	\caption{Optimal region for each case.}\label{single_block_fig2eps}
\end{figure}

Then, we investigate the performance for different values of $e^{\rm lim}$ and $e^{\rm avg}$. The numerical results are shown in Fig. \ref{single_block_fig1_degree1eps} and Fig. \ref{single_block_fig1_degree2eps}, from different angles. The gray dashed mesh is the globally optimal solution obtained by directly numerically solving (P1). The red surface corresponds to case (a) in Theorem \ref{theorem2}, meaning that we make a trade-off between transmitting power for information transmission and energy harvesting. It is obtained by solving \eqref{Casea}. Similarly, the green and the blue surfaces correspond to maximizing transmitting power for information transmission (case (b) in Theorem \ref{theorem2}) and maximizing transmitting power for energy harvesting (case (c) in Theorem \ref{theorem2}), which are obtained by solving \eqref{Caseb} and \eqref{Casec}, respectively. We can see the gray dashed mesh beautifully covers the other surfaces. It validates Algorithm \ref{al1}, which solves (P1) by choosing the best one from solutions of \eqref{Casea}, \eqref{Caseb}, and \eqref{Casec}. Then, in Fig. \ref{single_block_fig2eps}, we show the optimal region for each case. When $e^{\rm lim}$ and $e^{\rm avg}$ fall in the red region, the ($e^{\rm I}$, $\theta$) obtained by case (a) in Theorem \ref{theorem2} has better performance than other two cases. Similarly, ($e^{\rm I}$, $\theta$) obtained by case (b) is optimal in green region and the one obtained by case (c) is optimal in blue region. The results shows that for different average and peak power constraints, the optimal transmission schemes are also different. In this example, when there is a relatively strict constraint on peak power, we should maximize the transmitting power for information transmission. When both average power constraint and peak power constraint are loose, we should maximize the transmitting power for energy harvesting. Then, for other situations, we should perform a trade-off between them.

\subsection{Multiple Blocks}
\begin{figure}[!h]
	\begin{center}
		\hspace*{-10mm}
		\includegraphics[scale=0.8]{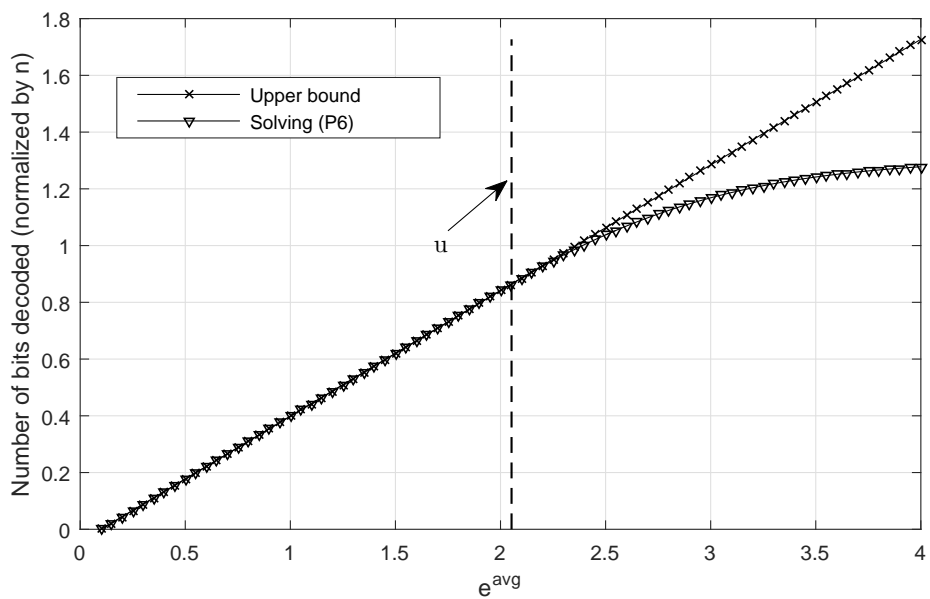}
	\end{center}
	\caption{Upper bounds and results directly solving (P6) for different values fo $e^{\rm avg}$.}\label{multiple_blocks_fig1eps}
\end{figure}

In this subsection, we investigate the performance over multiple blocks. We assume there are a total of 4 transmission blocks.  We set $g_1=g_2=g_3=g_4=g$, then \eqref{theorem3_equality} becomes $\dot{G}\leq g$. By the definition of $\dot{G}$, we can obtain that
\begin{eqnarray}
u = \frac{e^{\rm lim}-\dot{e}^{\rm I}}{\eta e^{\rm lim}+\mathcal{E}_{\rm D}(\dot{\theta})}\cdot g +\frac{\eta \dot{e}^{\rm I}+\mathcal{E}_{\rm D}(\dot{\theta})}{\eta \dot{e}^{\rm lim}+\mathcal{E}_{\rm D}(\dot{\theta})}\cdot e^{\rm lim},
\end{eqnarray}
and when $e^{\rm avg}\leq u$, the upper bound provided by Theorem \ref{theorem3} can be achieved; otherwise, the upper bound cannot be achieved. To illustrate this, we set $e^{\rm lim} = 4$, $g = 0.1$, and $\eta =1$, and plot the upper bounds and results directly solving (P6) for different values of $e^{\rm avg}$, respectively. From Fig. \ref{multiple_blocks_fig1eps}, we can see that to the left of the dashed line, the two curves overlap, and to the right of dashed line, a gap appears. This coincides with our analysis in Theorem \ref{theorem3}.

\section{Conclusions}
In this paper, we consider an end-to-end communication with an energy harvesting receiver. The transmitter works as a dedicated energy source and transfers some energy to the receiver, which can also harvest energy from ambient sources. When there are both average and peak power constraints at the transmitter, we maximize the total number of bits decoded at the receiver by optimizing the power used for energy harvesting, power used for information transmission, fraction of time for energy harvesting, and code rate. A generalized function is used to characterize the energy consumed at the receiver. For the single-block case, we provide an algorithm to obtain the global optimal solution. For the multiple-block case, we provide an upper bound and show when this bound can be achieved. In addition, an iterative method is given to get local optimal solution for the general case. Finally, we give some numerical results to illustrate our results and analysis.
\begin{appendices}
\section{}
Now we prove $\frac{\partial^2 C(e^{\rm I})}{\partial e^{\rm I2}}\leq 0$. Firstly, we can obtain that
\begin{eqnarray}
\frac{\partial^2 C(e^{\rm I})}{\partial e^{\rm I2}}=\frac{e^{-e^{\rm I}}(e^{\rm I})^{-\frac{3}{2}}}{\sqrt{4\pi}\ln 2}\cdot \phi(e^{\rm I}),
\end{eqnarray}
where
\begin{eqnarray}
\phi(e^{\rm I}) = (\ln\epsilon -\ln (1-\epsilon))(e^{\rm I}+0.5)+\Big(\frac{1}{\epsilon}+\frac{1}{1-\epsilon}\Big)\frac{e^{-e^{\rm I}}(e^{\rm I})^{\frac{1}{2}}}{\sqrt{4\pi}},
\end{eqnarray}
and $\epsilon=Q(\sqrt{2e^{\rm I}})$.
Now we only need to show $\phi(e^{\rm I})\leq 0$ when $e^{\rm I} \geq 0$. It is easy to obtain that
\begin{eqnarray}
\frac{\partial \phi(e^{\rm I})}{\partial e^{\rm I}}=\ln \epsilon -\ln(1-\epsilon)-\Big(\frac{1}{\epsilon}+\frac{1}{1-\epsilon}\Big)\frac{e^{-e^{\rm I}}(e^{\rm I})^{\frac{1}{2}}}{{4\pi}\epsilon(1-\epsilon)}\cdot\psi(e^{\rm I}),
\end{eqnarray}
where 
\begin{eqnarray}
\psi(e^{\rm I})=4\sqrt{\pi}(\epsilon-\epsilon^2)-(1-2\epsilon){e^{-e^{\rm I}}(e^{\rm I})^{-\frac{1}{2}}}.
\end{eqnarray}

If we can show $\psi(e^{\rm I})\geq 0$ when $e^{\rm I} \geq 0$, then $\frac{\partial^2 C(e^{\rm I})}{\partial e^{\rm I2}}\leq 0$ must hold. We have
\begin{eqnarray}
\frac{\partial \psi(e^{\rm I})}{\partial e^{\rm I}}=e^{-e^{\rm I}}(e^{\rm I})^{-\frac{3}{2}}\chi(e^{\rm I}),
\end{eqnarray}
where $\chi(e^{\rm I})=-(1-2\epsilon)e^{\rm I}-\frac{2}{\sqrt{4\pi}}e^{-e^{\rm I}}(e^{\rm I})^{\frac{1}{2}}+\frac{1}{2}-\epsilon$. It is easy to obtain that $\chi(e^{\rm I})\leq0$ when $e^{\rm I}\geq 0$, so $\psi(e^{\rm I})$ is non-increasing. We can see $\psi(e^{\rm I})\rightarrow 0$ as $e^{\rm I}\rightarrow +\infty$, so $\psi(e^{\rm I})\geq 0$ when $e^{\rm I} \geq 0$. Thus we prove $\frac{\partial^2 C(e^{\rm I})}{\partial e^{\rm I2}}\leq 0$.

\end{appendices}

\bibliographystyle{IEEEtran}
\bibliography{IEEEabrv,mybibfile}

\end{document}